\documentclass[aps,prd,nofootinbib,groupedaddress]{revtex4}
\usepackage{graphicx}
\usepackage{epsfig}
\usepackage{dcolumn}
\usepackage{amsfonts}
\usepackage{amssymb}
\usepackage{bm}
\usepackage{amsmath}
\def\Journal#1#2#3#4{{#1} {\bf#2}, #3 (#4)}

\def\NPA{{\rm Nucl. Phys.} A}
\def\NPB{{\rm Nucl. Phys.} B}
\def\PLB{{\rm Phys. Lett.}  B}

\def\PRD{{\rm Phys. Rev.} D}

\def\ep{\epsilon}

\def\la{\langle}
\def\ra{\rangle}

\def\lam{\lambda}

\def\be{\begin{equation}}
\def\ee{\end{equation}}
\def\bea{\begin{eqnarray}}
\def\eea{\end{eqnarray}}
\begin{document}
\title{Self-consistent covariant description of vector meson decay constants and chirality-even quark-antiquark distribution amplitudes
up to twist 3 in the light-front quark model}
%
\author{Ho-Meoyng Choi\\
Department of Physics, Teachers College, Kyungpook National University, Daegu 702-701, Korea\\
Chueng-Ryong Ji \\
 Department of Physics, North Carolina State University,
Raleigh, NC 27695-8202, U.S.A.\\
}

\begin{abstract}
Although the meson decay amplitude described by a two-point function may be regarded as one of the simplest possible physical
observables,
it is interesting that this apparently simple amplitude bears abundant fundamental
information
on QCD vacuum dynamics and chiral symmetry.
The light-front zero-mode issue of the vector meson decay constant $f_V$ is in this respect highly non-trivial and deserves careful analyses.
We discuss the zero-mode issue in the light-front quark model (LFQM) prediction of $f_V$ from the perspective of the vacuum fluctuation
consistent with the chiral symmetry of QCD.
We extend the exactly solvable manifestly covariant Bethe-Salpeter model calculation to the more phenomenologically accessible realistic light-front quark model and present a self-consistent covariant description of $f_V$
analyzing the twist-2 and twist-3 quark-antiquark distribution amplitudes with even chirality.
\end{abstract}

\pacs{xxx, yyy, zzz}

\maketitle

\section{Introduction}
 Meson decay constants provide essential
information on the QCD interaction between a quark and an antiquark.
They are the lowest moments of the light-cone distribution amplitudes (DAs) for
a quark and an antiquark inside a meson.
They are also important ingredients
 in studying
 the
 CP violation in
 leptonic or nonleptonic weak decays of
 mesons.

 Many theoretical works have been devoted to predict these fundamental constants of
 mesons,
 e.g. the lattice QCD~\cite{LQCD},  the QCD sum rules~\cite{QCDSR} and
 the light-front quark model~(LFQM)~\cite{Jaus90,Jaus99,Jaus03,Cheng04,CJ_99,CJ_DA,Choi07,BPP,MF12,BCJ_spin1,CJ_Bc,CJ_PV,CJ05}
 based on the LF  quantization~\cite{BPP} of QCD.
Among various theoretical approaches, the LFQM  has been successful in computing not only
the meson mass spectra~\cite{CJ_99,CJ_Bc} but also the decay constants and
weak transition form factors of
mesons~\cite{Jaus90,Jaus99,Jaus03,Cheng04,CJ_99,CJ_DA,Choi07,BPP,MF12,BCJ_spin1,CJ_Bc,CJ_PV,CJ05}.
In particular, the light-front dynamics (LFD) carries the
maximum
number~ (seven) of the kinetic (or interaction independent) generators and thus the less effort in dynamics is necessary in order to get the QCD solutions that reflect the full Poincar$\acute{e}$ symmetries. Also, the rational energy-momentum dispersion relation of
LFD yields the sign correlation between the LF energy $k^-(=k^0-k^3)$ and the LF longitudinal momentum $k^+(= k^0 + k^3)$ and leads to the suppression of vacuum fluctuations.

Despite these advantages in LFD, the zero-mode~($k^+ = 0$)~\cite{Zero} complication in the
matrix element has been noticed for the vector meson decay constant $f_V$~\cite{Jaus99,Jaus03,Cheng04,BCJ_spin1,CJ_fv13} as
well as some electroweak form factors involving a spin-1 particle~\cite{Jaus99,Jaus03,Cheng04,BCJ_spin1,CJ_PV,MF12}.
For the case of $f_V$, there has been a debate about the zero-mode
contribution to the matrix element of the plus component of the weak current $J^\mu_W$.
Unlike the electroweak form factor described by a three-point function involving an external probe, the meson decay amplitude is described by a two-point function and may be regarded as one of the simplest possible physical
observables.
It is interesting that this apparently simple amplitude bears abundant fundamental
information
on QCD vacuum dynamics and chiral symmetry.
In this respect, the zero-mode issue of $f_V$ in LFD is highly non-trivial and deserves careful analyses.
Indeed, we found~\cite{BCJ_spin1,CJ_fv13} that the existence or absence
of the zero mode may depend on the model, especially on the form of vector meson vertex operator $\Gamma^\mu$,
while Jaus~\cite{Jaus99,Jaus03} claimed that there exists zero-mode contribution to
$f_V$ even for the case of the good current $J^+_W$.
The purpose of this work is not just to clarify this zero-mode issue in the $f_V$ prediction from LFQM but to discuss this
topic in relation to the vacuum fluctuation
consistent with the chiral symmetry of QCD.
With this aim, we attempt to extend our previous analysis~\cite{CJ_fv13} from the exactly solvable manifestly covariant Bethe-Salpeter (BS) model to the more phenomenologically accessible realistic LFQM~\cite{CJ_99,CJ_DA,Choi07,Jaus90,Cheng97,Hwang10,Kon}
and discuss a self-consistent covariant description of the vector meson decay constant in view of the link between QCD and LFQM.

A systematic study of twist-3 light-cone DAs
of vector mesons in QCD was presented in Refs.~\cite{Ball96,Ball98}.
It was based on
the
conformal expansion taking into account meson and quark mass corrections. Two-particle DAs of vector mesons were classified in the same way as the more familiar  nucleon structure functions, i.e. parton distribution functions (PDFs), which correspond to the independent tensor structures in nonlocal matrix elements.  Nowadays, we know that there are twelve independent generalized parton distributions in deeply virtual Compton scattering~\cite{JB13}.  For the forward case with zero skewness, however, nine independent PDFs are found and classified by twist, spin and chirality~\cite{Jaffe-Ji}.  Similarly, the analysis of vector meson DAs revealed an analogous pattern as the operator structures are the same
with the case of PDFs
and the $\rho$ meson polarization vector formally substitutes the nucleon spin vector in the Lorentz structures~\cite{Ball98}.  Eight independent two-particle DAs were found with the classification due to twist, spin and chirality~\cite{Ball98}.  In this work, we focus on the chirality even distributions up to twist 3.

The light-cone DAs of a vector meson are defined in terms of the
following matrix elements of quark-antiquark non-local gauge invariant operators at light-like
separation~\cite{Ball96,Ball98,BJ07,FS11}:
\be\label{DAap1}
\la 0|{\bar q}(0)[0,z]\gamma^\mu q(z)|V(P, h)\ra
= f_V M \int^1_0 dx e^{-ixP\cdot z}
\biggl\{
P^\mu \frac{\ep_h \cdot z}{P\cdot z} \phi^{||}_{2;V}(x)
+ \biggl( \ep^\mu_h - P^\mu \frac{\ep_h \cdot z}{P\cdot z}
\biggr) \phi^{\perp}_{3;V}(x) + (\cdots)z^\mu
\biggr\},
\ee
where $z^2=0$ and the path-ordered gauge factor
\be\label{DAap2}
[0, z] = {\rm P}\exp[ ig\int^1_0 dt (-z)_\mu A^\mu ( (1-t) z)]
\ee
ensures the gauge invariance of the matrix elements and is equal to unity in the LF
gauge $A^+=0$. According to the classification of Ball and Braun~\cite{Ball96,Ball98},
$\phi^{||}_{2;V}(x)$ and $\phi^{\perp}_{3;V}(x)$ correspond to the twist-2 and twist-3 two particle
DAs, respectively. The ellipses in Eq.~(\ref{DAap1}) represent the higher twist
contribution~\cite{Ball96,Ball98,BJ07,FS11}
which we do not consider in this work. The
normalization of the two DAs $\Phi=\{\phi^{||}_{2;V},\phi^{\perp}_{3;V} \}$ is given by
\be\label{DAap3}
\int^1_0 dx \; \Phi(x) = 1.
\ee
In order to establish a connection between these DAs and the LF wave functions of vector mesons, we
need to apply the equal LF time condition, $z^+=0$, and choose the LF gauge $A^+=0$.
Then, neglecting the higher twist DAs, Eq.~(\ref{DAap1}) can be rewritten as
\be\label{DAap4}
\la 0|{\bar q}(0)\gamma^\mu q(z)|V(P, h)\ra|_{z^+={\bf z}_\perp=0}
= f_V M \int^1_0 dx e^{-ixP\cdot  z}
\biggl\{
P^\mu \frac{\ep_h^+}{P^+} \phi^{||}_{2;V}(x)
+ \biggl( \ep^\mu_h - P^\mu \frac{\ep_h^+}{P^+}
\biggr) \phi^{\perp}_{3;V}(x)
\biggr\}.
\ee
To isolate the twist-2 DA, $\phi^{||}_{2;V}$, we may take the plus component ($\mu=+$) of the
current
with the longitudinal
polarization ($h=0$) and obtain
\be\label{DAap5}
\la 0|{\bar q}(0)\gamma^+ q(z^-)|V(P, h)\ra
= f_V M \ep_{0}^+ \int^1_0 dx e^{-ixP \cdot z}
\phi^{||}_{2;V}(x).
\ee
On the other hand, to isolate the twist-3 DA, $\phi^{\perp}_{3;V}$, we take the perpendicular
component
($\mu=\perp$) of the
current
with the transverse polarization ($h=+$) and obtain
\be\label{DAap6}
\la 0|{\bar q}(0)\gamma^\perp q(z^-)|V(P, h)\ra
= f_V M \ep_{+}^\perp \int^1_0 dx e^{-ixP \cdot z}
\phi^{\perp}_{3;V}(x).
\ee

It is a common practice to utilize an exactly solvable manifestly covariant model
to check the existence (or absence) of the zero-mode and substitute the radial and spin-orbit wave functions
of the exactly solvable model with the more phenomenologically accessible model wave functions that can be
provided by LFQM. To discuss the nature of the LF zero-mode in meson decay amplitude,
we may denote the total LF longitudinal momentum of the meson, $P^+ = k_Q^+ + k_{\bar Q}^+$, where
$k_Q^+$ and $k_{\bar Q}^+$ are the individual quark and antiquark LF longitudinal momenta, respectively.
Similarly, the LF energy $P^-$ is shared by $k_Q^-$ and $k_{\bar Q}^-$, i.e. $P^- = k_Q^- + k_{\bar Q}^-$.
The LF energy integration is done typically by using
the Cauchy's theorem for a contour integration.
For the LF energy integration of the two-point function to compute the meson decay amplitude,
one may pick up a LF energy poles, e.g. either $[k_Q^-]_{\rm on}$  (i.e. on shell value of $k_Q^-$)
from the quark propagator or $[k_{\bar Q}^-]_{\rm on}$ from the antiquark propagator.
However, it is crucial to note that the poles move to infinity (or fly away in the complex plane) as the LF longitudinal momentum,
either $k_Q^+$ or $k_{\bar Q}^+$, goes to zero~\cite{BDJM} .
Unless the contribution from the pole flown into infinity vanishes, it must be kept in computing the physical
observable. Since such contribution,
if it exists,
appears either from $k_Q^+=0$ and $k_{\bar Q}^+=P^+$ or
from $k_{\bar Q}^+=0$ and $k_Q^+=P^+$, we call it as the zero-mode contribution.
In our previous work~\cite{CJ_PV,CJ_Bc,CJ_fv13}, we discussed the power-counting method which can reveal the
existence or absence of the zero-mode contribution by analyzing the power behavior of the
integration variable such as $k_Q^+$ or $k_{\bar Q}^+$ and
presented
an
effective method of  identifying the corresponding zero mode operators.
If the zero-mode exists, it is critical to take into account its contribution in order to get the identical result to the one obtained by manifestly covariant calculation.

As discussed above, in the case of two-point function for the computation of the meson decay constant,
the zero-mode contribution is locked into a single point of the LF longitudinal momentum, i.e. either $k_Q^+=0$ where
$k_{\bar Q}^+=P^+$ or $k_{\bar Q}^+=0$ where $k_Q^+=P^+$.
Since one of the constituents of the meson carries the entire momentum $P^+$ of the meson,
the other constituent carries the zero LF longitudinal momentum and thus can be regarded as the zero-mode quantum
fluctuation linked to the vacuum. This link is due to
a pair creation of particles with zero LF longitudinal momenta from the vacuum.
It is important to capture the vacuum effect for the consistency with the chiral symmetry properties of the strong
interactions~\cite{JMT2013}.
With this link, the zero-mode contribution in the meson decay process can be considered effectively
as the effect of vacuum fluctuation consistent with the chiral symmetry of the strong interactions.
In this respect, the LFQM with effective degrees of freedom represented by the constituent quark and antiquark
may be linked to the QCD. The zero-mode link to the QCD vacuum may provide the view of effective zero-mode cloud
around the quark and antiquark inside the meson and the constituents dressed by the zero-mode cloud may satisfy the chiral symmetry
consistent with the QCD. Since the constituent quark and antiquark used in the LFQM have already absorbed the zero-mode cloud,
the zero-mode contribution in the LFQM may not be as explicit as in the manifestly covariant model calculation but
provide effectively the consistency with the chiral symmetry.

In the LFQM presented in~\cite{CJ_99,CJ_DA,Choi07,Jaus90,Cheng97,Hwang10,Kon}, the constituent quark and antiquark in a bound
state are required to be on-mass-shell, which is different from the covariant formalism in which the constituents
are off-mass-shell. The spin-orbit wave function in the LFQM is obtained by the interaction-independent Melosh
transformation~\cite{Melosh} from the ordinary equal-time static spin-orbit wave function assigned by the
quantum number $J^{PC}$. The common feature of the LFQM presented in~\cite{CJ_99,CJ_DA,Choi07,Jaus90,Cheng97,Hwang10,Kon}
is to use the sum of the light-front energy of the constituent quark and antiquark for the meson mass in the spin-orbit wave function.
In the standard light-front (SLF) approach used in
the LFQM~\cite{CJ_99,CJ_DA,Choi07,Jaus90,Cheng97,Hwang10,Kon},
the vector meson decay constant $f^{SLF}_V$ is obtained by the matrix element of the plus component of the
currents in 3-dimensional LF momentum space. As the constituent quark and antiquark in LFQM
are the dressed constituents including the zero-mode effect,
the SLF approach within the phenomenological LFQM~\cite{CJ_99,CJ_DA,Choi07,Jaus90,Cheng97,Hwang10,Kon}
is not amenable to determine the zero-mode contribution by itself. We thus utilize a manifestly covariant model
to check the existence (or absence) of the zero-mode and substitute the radial and spin-orbit wave functions
with the phenomenologically accessible model wave functions provided by the LFQM analysis of meson mass spectra.
If the on-mass-shell spin structure of the matrix element in SLF approach is not exactly reproducible from the manifestly covariant model, the SLF result is identified to be necessary to take into account  the zero-mode. We discuss the zero-mode operator necessary for the SLF analysis of the meson decay process.

To further clarify the zero-mode issue regarding on $f_V$, we analyze the twist-2 and twist-3 two-particle DAs and examine a
fundamental constraint anticipated from the LFQM~\cite{CJ_99,CJ_DA,Choi07,Jaus90,Cheng97,Hwang10,Kon}:
i.e. symmetric quark-antiquark DAs for the equal quark and antiquark bound state mesons such as $\rho$.
As we shall show in this work, the existence of zero-mode contribution to $f_V$ claimed by Jaus~\cite{Jaus99,Jaus03}
and subsequently advocated by other authors~\cite{Cheng04} contradicts with this anticipated constraint.
We also note that the two equivalent decay constants obtained from $(J^\mu_W,\epsilon_h)=(J^+_W,\epsilon_{0})$
and $(J^\perp_W,\epsilon_{+})$ are related to the twist-2 and twist-3 two-particle DAs of
a vector meson~\cite{Ball96,Ball98,BJ07,FS11}, respectively.

The paper is organized as follows: In Sec.~\ref{sec:II}, we briefly discuss the vector
meson decay constant in an exactly solvable model based on the covariant BS model of
(3+1)-dimensional fermion field theory. We then present our LF calculation
of the vector meson decay constant using two different combinations of
LF weak currents $J^\mu_W$ and polarization vectors $\epsilon_h$, i.e.
$(J^+_W,\epsilon_{0})$
and $(J^\perp_W,\epsilon_{+})$ and check the LF covariance of the decay constant within
the covariant BS model. Especially, we identify the zero-mode contributions to the decay constant
and find the corresponding zero-mode operators.
In Sec.~\ref{sec:III}, we present the SLF calculation of the decay constant in a phenomenologically
more realistic LFQM with the gaussian wave function.
In Sec.~\ref{sec:cons}, we discuss the correct relation linking the manifestly covariant model to the standard LFQM.
We present self-consistent covariant descriptions of vector meson decay constants as well as
twist-2 and twist-3 two-particle DAs in the standard LFQM.
In Sec.~\ref{sec:Num}, we present our numerical results for the explicit demonstration
of our findings. Summary and discussion follow in Sec.~\ref{sec:sum}. The details of the
spin structure in standard LFQM are summarized in Appendix~\ref{sec:spin} and the analyses
of the pseudoscalar meson decay constant are presented in Appendix~\ref{sec:psfv}.

\section{Manifestly Covariant Model}
\label{sec:II}
The decay constant $f_V$ of a vector meson with the four-momentum $P$ and the mass $M$
as a $q{\bar q}$ bound state is defined by the matrix element of the vector
current

\be\label{eq:1}
\la 0|{\bar q}\gamma^\mu q|V(P,h)\ra
= f_V M \epsilon^\mu_h,
\ee
where the polarization vector $\epsilon_h$ of a vector meson satisfies the Lorentz condition
$\epsilon_h \cdot P = 0$.

\begin{figure}
\begin{center}
\includegraphics[height=2.5cm, width=5cm]{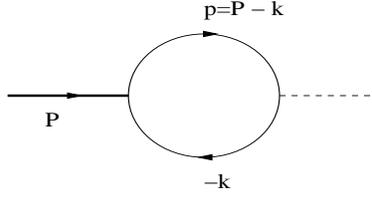}
\caption{\label{fig1} Feynman diagram for a decay constant. }
\end{center}
\end{figure}
The matrix element $A^\mu_h \equiv \la 0|{\bar q}\gamma^\mu q|V(P,h)\ra$ is given in
the one-loop approximation~(see Fig.~\ref{fig1}) as a momentum integral
\be\label{eq:2}
A^\mu_h = N_c
\int\frac{d^4k}{(2\pi)^4} \frac{H_V} {N_p N_k} S^\mu_h,
\ee
where $N_c$ denotes the number of colors. The denominators
$N_p$ and $N_k$ come from the fermion propagators
of mass $m_{1}$ and $m_2$ carrying the internal four-momenta $p =P -k$ and $k$, respectively,
and they are given by $N_p  = p^2 -m^2_1 +i\varepsilon$ and
$N_k = k^2 - m^2_2+i\varepsilon$.
In order to regularize the covariant loop in $(3+1)$ dimensions,
we use the usual multipole ansatz~\cite{Jaus99} for the $q{\bar q}$ bound-state vertex function
$H_V$ of a vector meson:
\be\label{eq:5}
H_V = \frac{g}{N_\Lambda^n},
\ee
where $N_\Lambda  = p^2 - \Lambda^2 +i\varepsilon$, and $g$ and $\Lambda$ are constant parameters
and the power $n$ for the multipole ansatz should
be $n\geq 2$ for the regularization.
%
For our purpose,
we take $n=2$ since our qualitative results in conjunction with the zero-mode issue
do not depend on the value of $n$.

The trace term $S^\mu_h$  in Eq.~(\ref{eq:2}) is given by
\be\label{eq:3}
 S^\mu_h  =  {\rm Tr}\left[\gamma^\mu\left(\slash \!\!\!p+m_1 \right)
 \Gamma\cdot\epsilon_h
 \left(-\slash \!\!\!k + m _2\right) \right],
\ee
where the vector meson vertex operator $\Gamma^\mu$
is given by
\be\label{eq:4}
\Gamma^\mu =\gamma^\mu-\frac{(p-k)^\mu}{D}.
\ee
While the Dirac coupling $\gamma^\mu$ is intrinsic to the vector meson vertex, the model-dependence of a
vector meson is implemented through the factor $D$ in Eq.~(\ref{eq:4}).
For the explicit comparison between the manifestly covariant calculation and the LF calculation,
checking the existence (or absence) of the zero-mode contribution to the vector meson decay constant,
we analyze $\Gamma^\mu$ in this section with a constant $D$ factor,
i.e. $D=D_{\rm con}=M + m_1 + m_2$.
We will discuss the more realistic (but not manifestly covariant) $D$ factor such as $D_{\rm LF}=M_0 + m_1 + m_2$ with
the invariant mass $M_0$ of the vector meson in~Sec.\ref{sec:III} devoted to phenomenologically more
accessible LFQM~\cite{CJ_99,CJ_DA,Choi07,Jaus90,Cheng97,Hwang10,Kon}.

The manifestly covariant result for $n=2$ case is given by~\cite{CJ_fv13}
\bea\label{eq:7}
f_V^{\rm cov} &=& \frac{N_c g}{4\pi^2 M} \int^1_0 dx\int^{1-x}_0 dy (1-x-y)
\biggl\{ \frac{y(1-y)M^2 + m_1 m_2}{C^2_{\rm cov}}
- \frac{1 + \frac{m_1 + m_2}{D_{\rm con}}}{C_{\rm cov}} \biggr\},
\eea
where
\be\label{eq:8}
C_{\rm cov} = y(1-y) M^2 - x m^2_1 - y m^2_2 - (1-x-y) \Lambda^2.
\ee


Performing the LF calculation in parallel with the
manifestly covariant one, we use two different combinations of the currents and the polarization vectors,
i.e. (1) plus component~($\mu=+$) of the currents with the longitudinal
polarization $\epsilon^\mu_{0}$ and (2) perpendicular components ($\mu=\perp$)
of the currents with the transverse polarization $\epsilon^\mu_{\pm}$,
to obtain the decay constant.
The polarization vectors used in this manifestly covariant analysis are given by
\bea\label{eq:9}
\epsilon^\mu_0&=& [\ep^+,\ep^-,\ep_\perp] =
\frac{1}{M}\biggl[P^+,\frac{{\bf P}^2_{\perp}-M^2}{P^+},
{\bf P}_{\perp}\biggr],\nonumber\\
\epsilon^\mu_{\pm} &=&
\biggl[0,\frac{2}{P^+}{\bf\epsilon}^\perp_{\pm}\cdot{\bf P_{\perp}},
{\bf\epsilon}^\perp_{\pm}\biggr],\;
{\bf\epsilon}^\perp_{\pm} = \mp\frac{(1,\pm i)}{\sqrt{2}},
\eea
where we use the metric convention
$a\cdot b =\frac{1}{2}(a^+b^- + a^-b^+)-{\bf a}_\perp\cdot{\bf b}_{\perp}$.

The trace term $S^\mu_h$ in Eq.~(\ref{eq:3}) can be separated into the
on-mass-shell propagating part $[S^\mu_h]_{\rm on}$ and
the off-mass-shell instantaneous part $[S^\mu_h]_{\rm inst}$ via
$\slash \!\!\!q = \slash \!\!\!q_{\rm on}  + \frac{1}{2}\gamma^+(q^- - q^-_{\rm on})$
as
\be\label{eq:10}
S^\mu_h = [S^\mu_h]_{\rm on} + [S^\mu_h]_{\rm inst},
\ee
where
\bea\label{eq:11}
[S^\mu_h]_{\rm on} =
4 [\epsilon^\mu_h (p_{\rm on}\cdot k_{\rm on} + m_1 m_2)
- p^\mu_{\rm on}(\epsilon_h \cdot k_{\rm on})
-k^\mu_{\rm on}(\epsilon_h \cdot p_{\rm on}) ]
+ 8\frac{\epsilon_h \cdot k_{\rm on}}{D} (m_2 p^\mu_{\rm on} - m_1 k^\mu_{\rm on}),
\eea
and
\bea\label{eq:12}
[S^\mu_h]_{\rm inst} = 2\Delta_{k}^- \left(\epsilon^\mu_h  p^+ - \epsilon^+_h p^\mu_{\rm on}
 \right)
+ 2 \Delta_{p}^- \left(\epsilon^\mu_h k^+ - \epsilon^+_h k^\mu_{\rm on}
 \right)
+ 4 \frac{\epsilon^+_h \Delta_{k}^-}{D} (m_2 p^\mu_{\rm on} - m_1 k^\mu_{\rm on})
+ (\cdots)g^{\mu +},
\eea
with $\Delta_{q}^\mu=q^\mu -q_{\rm on}^\mu$. Since the $g^{\mu+}$ term in Eq.~(\ref{eq:12})
vanishes when $\mu=+$ or $\perp$ is taken, we do not list the explicit forms of the ellipses
in front of $g^{\mu+}$.
Furthermore, we take the reference frame where ${\bf P}_\perp =0$, i.e.,
$P=( P^+, M^2/P^+, 0)$. In this case, the four momenta of the on-mass-shell
constituents are given by
\bea\label{eq:13}
 p_{\rm on} &=& \biggl[ x P^+, \frac{{\bf k}^2_\perp + m^2_1}{x P^+}, -{\bf k}_\perp \biggr],
 \nonumber\\
 k_{\rm on} &=& \biggl[ (1-x) P^+, \frac{{\bf k}^2_\perp + m^2_2}{(1-x) P^+}, {\bf k}_\perp \biggr],
\eea
where $x=p^+/P^+$ is the LF longitudinal momentum fraction of the quark.

By the integration over $k^-$ in Eq.~(\ref{eq:2}) and closing the contour in the lower
half of the complex $k^-$ plane, one picks up the residue at $k^-=k^-_{\rm on}$
 in the region $0< k^+ < P^+$ (or $0 < x < 1$).
 Thus, the Cauchy integration formula for the $k^-$
integral in Eq.~(\ref{eq:2}) yields
\be\label{eq:14}
 A^\mu_h = \frac{N_c}{16\pi^3}\int^{1}_0
 \frac{dx}{(1-x)} \int d^2{\bf k}_\perp
 \chi(x,{\bf k}_\perp) S^\mu_h(k^-=k^-_{\rm on}),
\ee
where
\be\label{eq:15}
\chi(x,{\bf k}_\perp) = \frac{g}{[x (M^2 -M^2_0)][x (M^2 - M^2_{\Lambda})]^n},
\ee
and
\bea
\label{eq:16}
 M^2_{0(\Lambda)} &=& \frac{ {\bf k}^{2}_\perp + m_1^2(\Lambda^2)}{x}
 + \frac{ {\bf k}^{2}_\perp + m^2_2}{1-x}.
\eea
Note that the second term in the denominator of Eq.~(\ref{eq:15}) comes from the multipole type
vertex function $H_V$ defined in Eq.~(\ref{eq:3}).
Although we take $n=2$ for a direct comparison with the manifestly covariant result $f^{\rm cov}_V$,
the qualitative result regarding on the zero-mode issue is independent of the power  $n~(\geq 2)$ as we shall show.

\subsection{ Decay constant from longitudinal polarization}

Using the plus component ($\mu=+$) of the currents with the longitudinal
polarization vector $\epsilon^\mu_{0}$ in Eq.~(\ref{eq:9}),
the decay constant is obtained from the relation
\be\label{eq:17}
f^{(h=0)}_V = \frac{A^+_{h=0}}{M \epsilon^+_0}.
\ee
For the purpose of analyzing zero-mode contribution to the decay constant,
we denote the decay constant as $[f^{(h=0)}_V]_{\rm val}$ (meaning the valence
contribution to the decay constant)
when the matrix element $A^+_{0}$ is obtained for $k^-=k^-_{\rm on}$ in the
region of $0<x<1$. Explicitly, it is given by
\bea\label{eq:18a}
[f^{(h=0)}_V]_{\rm val}
 &=& \frac{N_c}{4  \pi^3}\int^{1}_0
 \frac{dx}{(1-x)} \int d^2{\bf k}_\perp
 \chi(x,{\bf k}_\perp)
\frac{1}{M}\biggl\{ x(1-x)M^2 + {\bf k}^2_\perp + m_1 m_2
 \nonumber\\
 &&
+ [m_2 x - (1-x) m_1]
 \frac{ \left[ {\bf k}^2_\perp + m^2_2 - (1-x)^2 M^2 \right]}{(1-x)D_{\rm con}}
 \biggr\}.
\eea
We note that the valence contribution to the trace term in Eq.~(\ref{eq:14}) comes only
from the on-shell propagating part but not from the instantaneous one, i.e.
$[S^+_{0}]_{\rm val} = [S^{+}_{0}]_{\rm on}$  and
$[f^{(h=0)}_V]_{\rm val}=[f^{(h=0)}_V]_{\rm on}$.

Comparing $[f^{(h=0)}_V]_{\rm val}$ with the manifestly
covariant result $f^{\rm cov}_V$, we find that $[f^{(h=0)}_V]_{\rm val}$  is
exactly the same as $f^{\rm cov}_V$ for the model-independent Dirac coupling,
$\Gamma^\mu=\gamma^\mu$ (or $1/D=0$).
The same observation has also been made in Ref.~\cite{BCJ_spin1}.
However, $[f^{(h=0)}_V]_{\rm val}$ is different from $f^{\rm cov}_V$ when the model-dependent
$D=D_{\rm con}$ term is included. In this case, the difference between the two results,
$f^{\rm cov}_V - [f^{(h=0)}_V]_{\rm val}$,
corresponds to the zero-mode contribution $[f^{(h=0)}_V]_{\rm Z.M.}$ to the full solution
$[f^{(h=0)}_V]_{\rm full}=[f^{(h=0)}_V]_{\rm val} + [f^{(h=0)}_V]_{\rm Z.M.}$.
For the case of $D=D_{\rm con}$,
%
the zero-mode contribution to $f^{(h=0)}_V$ comes
from the singular $p^-$ (or equivalently $1/x$) term in $S^+_{0}$ in the limit
of $x\to 0$ when $p^-=p^-_{\rm on}$, i.e.
\be \label{eq:18}
\lim_{x\to 0}S^{+}_{0}(p^-=p^-_{\rm on})
= 4m_1\frac{\epsilon^+_{0} p^-}{D_{\rm con}}.
\ee
We note that the singular
term in Eq.~(\ref{eq:18}) comes only from the instantaneous contribution.

The necessary prescription to identify zero-mode operator corresponding to
$p^-$ is analogous to that derived in the previous analyses of weak transition form factor
calculations~\cite{Jaus99,CJ_Bc,CJ_PV}, except that there is no momentum transfer $q$ dependence.
By replacing $p^-$ with $-Z_2$~\cite{Jaus99,CJ_Bc,CJ_PV} in Eq.~(\ref{eq:18}),
we now identify the zero-mode operator $[S^{+}_{0}]_{\rm Z.M.}$
corresponding to Eq.~(\ref{eq:18}) as follows
\be\label{eq:19}
[S^{+}_{0}]_{\rm Z.M.} = 4m_1\frac{\epsilon^+_{0} (-Z_2)}{D_{\rm con}},
\ee
where
 \be\label{eq:20}
Z_2 = x(M^2 - M^2_0) + m^2_1 - m^2_2 + (1-2x)M^2.
 \ee
 This zero-mode operator $[S^{+}_{0}]_{\rm Z.M.}$ can be effectively included
in the valence region, i.e. the full (exact) solution of the trace term is given by
 $ [S^+_{0}]_{\rm full}= [S^{+}_{0}]_{\rm val} + [S^{+}_{0}]_{\rm Z.M.}$.  Or, equivalently,
 the zero-mode contribution to the decay constant,
$[f^{(h=0)}_V]_{\rm Z.M.} = [A^+_{0}]_{\rm Z.M.}/(M \epsilon^+_{0})$, is given by
\bea\label{eq:zm}
 [f^{(h=0)}_V]_{\rm Z.M.} &=& \frac{N_c}{16 \pi^3}\int^{1}_0
 \frac{dx}{(1-x)} \int d^2{\bf k}_\perp
 \chi(x,{\bf k}_\perp)
 \frac{ [S^{+}_{0}]_{\rm Z.M.} }{M \epsilon^+_0}.
\eea
%
Adding Eqs.~(\ref{eq:18a}) and~(\ref{eq:zm}),
we finally obtain the full result
of the decay constant for the longitudinal polarization as
\bea\label{eq:21}
 [f^{(h=0)}_V]_{\rm full}
 &=& \frac{N_c}{4  \pi^3}\int^{1}_0
 \frac{dx}{(1-x)} \int d^2{\bf k}_\perp
 \chi(x,{\bf k}_\perp)
\frac{1}{M} \biggl\{ x(1-x)M^2 + {\bf k}^2_\perp + m_1 m_2
 \nonumber\\
 &&+ x(m_1 + m_2)
 \frac{\left[ {\bf k}^2_\perp + m^2_2 - (1-x)^2 M^2 \right]}{(1-x)D_{\rm con}}
 \biggr\}.
\eea
It can be checked that
Eq.~(\ref{eq:21}) is identical to the manifestly covariant result of
Eq.~(\ref{eq:7}).

\subsection{ Decay constant from transverse polarization}
Secondly, using the perpendicular components ($\mu=\perp$) of the currents with the transverse polarization vector $\epsilon^\mu_+$,
the decay constant is obtained from the relation
\be\label{eq:22}
f^{(h=1)}_V = \frac{{\bf A}^\perp_{h=1} \cdot \epsilon^{\perp *}_+}{M}.
\ee
In this case, the valence contributions  to the trace term in Eq.~(\ref{eq:14}) come not only
from the on-shell part but also from the instantaneous one, i.e.
$[S^\perp_{+}]_{\rm val} = [S^{\perp}_{+}]_{\rm on}
+ 2 k^+ \epsilon^\perp_{+}\Delta_p^-$ where the latter corresponds to the
instantaneous contribution.
The on-shell contribution to the decay constant $f^{(h=1)}_V$ is given by
\bea\label{eq:23on}
 [f^{(h=1)}_V]_{\rm on} &=& \frac{N_c}{4 \pi^3}\int^{1}_0
 \frac{dx}{(1-x)} \int d^2{\bf k}_\perp
 \chi(x,{\bf k}_\perp)
\frac{1}{M}\biggl\{ \frac{{\bf k}^2_\perp + {\cal A}^2}{2x(1-x)} -{\bf k}^2_\perp
  + \frac{ (m_1 + m_2)}{D_{\rm con}}{\bf k}^2_\perp
 \biggr\},
\eea
where ${\cal A}= (1-x) m_1 + x m_2$
and the valence contribution $[f^{(h=1)}_V]_{\rm val}(=[f^{(h=1)}_V]_{\rm on} + [f^{(h=1)}_V]_{\rm inst})$
is given by
\bea\label{eq:23a}
 [f^{(h=1)}_V]_{\rm val} &=& \frac{N_c}{8\pi^3}\int^{1}_0
 \frac{dx}{(1-x)^2} \int d^2{\bf k}_\perp
 \chi(x,{\bf k}_\perp)
\frac{1}{M}\biggl\{ (2x -1) ({\bf k}^2_\perp + m^2_2) + 2 (1-x) m_1 m_2
 \nonumber\\
 &+& (1-x)^2 M^2 + 2 (1-x) \frac{ (m_1 + m_2)}{D_{\rm con}}{\bf k}^2_\perp
 \biggr\}.
\eea
We find that Eq.~(\ref{eq:23a}) does not coincide with the manifestly covariant result.
That is, the decay constant obtained from the perpendicular components of the currents with the
transverse polarization receives a zero mode.

Again, the zero-mode contribution to $f^{(h=1)}_V$  can be obtained from all possible singular
$p^-$ terms in $S^\perp_+$ in the limit of $x\to 0$ when $p^-=p^-_{\rm on}$.
From Eqs.~(\ref{eq:11}) and~(\ref{eq:12}), we find the nonvanishing
singular term as follows
\be \label{eq:23}
\lim_{x\to 0}S^{\perp}_+ (p^-=p^-_{\rm on})
= 2p^-\epsilon^\perp_+.
\ee
We should note that the singular term in Eq.~(\ref{eq:23}) comes only from the on-shell part
$[S^{\perp}_+]_{\rm on}$, but not from the instantaneous part $[S^{\perp}_+]_{\rm inst}$.
This implies that the zero-mode contribution to $f^{(h=1)}_V$ comes only from the model independent Dirac
coupling part, $\Gamma^\mu=\gamma^\mu$.
By the replacement $p^-\to -Z_2$ in Eq.~(\ref{eq:23})
as previously discussed in the derivation of Eq.~(\ref{eq:19}) from Eq.~(\ref{eq:18}),
we now obtain the corresponding zero-mode operator $[S^{\perp}_+]_{\rm Z.M.}$ as
\be \label{eq:24}
[S^{\perp}_+]_{\rm Z.M.}
= 2(-Z_2)\epsilon^\perp_+.
\ee

That is, the full solution of the trace term in the valence region
is $[S^\perp_+]_{\rm full}= [S^{\perp}_+]_{\rm val} + [S^{\perp}_+]_{\rm Z.M.}$.
Or the zero-mode contribution to the decay constant,
$[f^{(h=1)}_V]_{\rm Z.M.} = [{\bf A}^\perp_{h=1}]_{\rm Z.M.}\cdot\ep^{\perp *}_+/ M$,
is given by
\bea\label{eq:zmT}
 [f^{(h=1)}_V]_{\rm Z.M.} &=& \frac{N_c}{16 \pi^3}\int^{1}_0
 \frac{dx}{(1-x)} \int d^2{\bf k}_\perp
 \chi(x,{\bf k}_\perp)
 \frac{ [S^{\perp}_+]_{\rm Z.M.}\cdot\ep^{\perp *}_+ }{M}.
\eea

Adding Eqs.~(\ref{eq:23a}) and~(\ref{eq:zmT}), we now obtain the full result
of the decay constant for the transverse polarization as
\bea\label{eq:25}
 [f^{(h=1)}_V]_{\rm full} &=& \frac{N_c}{4 \pi^3}\int^{1}_0
 \frac{dx}{(1-x)} \int d^2{\bf k}_\perp
 \chi(x,{\bf k}_\perp)
\frac{1}{M}\biggl\{ x M^2_0 - m_1(m_1-m_2) -{\bf k}^2_\perp
  + \frac{ (m_1 + m_2)}{D_{\rm con}}{\bf k}^2_\perp
 \biggr\}.
\eea
One can check that
$[f^{(h=1)}_V]_{\rm full}$ is the same as $[f^{(h=0)}_V]_{\rm full}$ [Eq.~(\ref{eq:21})] as well as the
manifestly covariant result $f^{\rm cov}_V$.
We also confirm that our $[f^{(h=1)}_V]_{\rm full}$ is exactly the same as the one
obtained by Jaus~\cite{Jaus99} (see Eq.~(4.22) of Ref.~\cite{Jaus99}).

\section{Standard Light-Front Quark Model}
\label{sec:III}
In the standard LFQM~\cite{CJ_99,CJ_DA,Choi07,Jaus90,Cheng97,Hwang10,Kon}, the momentum space meson
wave function is given by
\be\label{ap0}
\Psi^{SS_z}_{\lam_1\lam_2}(x,{\bf k}_{\perp})
={\cal R}^{SS_z}_{\lam_1\lam_2}(x,{\bf k}_{\perp})
\phi(x,{\bf k}_{\perp}),
\ee
where $\phi$ is the radial
wave function and ${\cal R}^{SS_z}_{\lam_1\lam_2}$ is the
spin-orbit wave function that is obtained by the interaction-
independent Melosh transformation~\cite{Melosh} from the ordinary
spin-orbit wave function assigned by the quantum numbers
$J^{PC}$. The explicit form of the spin-orbit wave function of definite spin
$(S,S_z)$ is constructed out of LF helicity $(\lam_1,\lam_2)$ as follows
\bea\label{ap1}
{\cal R}^{SS_{z}}_{\lam_{1}\lam_{2}}(x,{\bf k}_{\perp})
\hspace{-0.2cm} &&=
\sum_{s_1,s_2}
\langle\lam_1 |{\cal R}^{\dagger}_{{\rm M}}(1-x,{\bf k}_{\perp},m_1)|s_1\rangle
\langle\lam_2 |{\cal R}^{\dagger}_{{\rm M}}(x,-{\bf k}_{\perp},m_2)|s_2\rangle
\langle\frac{1}{2}s_1\frac{1}{2}s_2|S S_z\rangle,
\eea
where $|s_i\ra$ are the usual Pauli spinor, and ${\cal R}_{{\rm M}}$ is the
Melosh transformation operator:
\be\label{ap2}
{\cal R}_{{\rm M}}(x_i,{\bf k}_{\perp},m_i)=
\frac{m_i + x_i M_{0} - i\sigma\cdot(\hat{\bf n}
\times {\bf k}_\perp)}{\sqrt{(m_i +x_i M_{0})^{2} + {\bf k}^{2}_{\perp}}},
\ee
with $\hat{\bf n}$=(0,0,1) being a unit vector in the $z$ direction.
The spin-orbit wave functions can also be represented in the
following covariant way:
\be\label{ap3}
{\cal R}_{\lam_1 \lam_2}^{SS_z}(x,{\bf k}_{\perp})
=\frac{\bar{u}_{\lam_1}(p_1)\Gamma v_{\lam_2}( p_2)}
{\sqrt{2}[M^{2}_{0}-(m_1-m_2)^{2}]^{1/2}}.
\ee
The vertex operator for a pseudoscalar meson is $\Gamma =\gamma_5$ and that for a vector meson
is given by~\cite{CJ_DA,Choi07,Jaus90,Cheng97,Hwang10,Kon}
\be\label{ap4}
\Gamma = -/\!\!\!\tilde{\ep} +\frac{\tilde{\ep}\cdot(p_1-p_2)}{D_{\rm LF}},
\ee
where $D_{\rm LF}=M_0+m_1+m_2$ and $\tilde{\ep}$ is the polarization vector
specified in the center of mass frame of the quark and
antiquark. We should note that
while the transverse polarization vector $\tilde{\ep}^\mu_{\pm}$ of the $q\bar{q}$ system
coincides with that of the meson polarization vector $\ep^\mu_{\pm}$ given by Eq.~(\ref{eq:9}),
the longitudinal polarization vector $\tilde{\ep}^\mu_0$ is different from $\ep^\mu_0$ and
is given by~\cite{CJ_DA,Choi07,Jaus90,Cheng97,Hwang10,Kon}
\be\label{ap5}
\tilde{\ep}^{\mu}_{0}=\frac{1}{M_{0}}\biggl[
P^+,\frac{-M^{2}_{0}+{\bf P}^2_{\perp}}{P^{+}},{\bf P}_{\perp}
\biggr].
\ee
That is, the invariant mass $M_0$ instead of the physical mass $M$
is used to define the vector meson vertex operator $\Gamma$.
We also should note that $\tilde{\ep}^\mu_0$ is used only in the vector meson vertex operator in Eq.~(\ref{ap4}).
The virtue of using $M_0$
is to satisfy the normalization of ${\cal R}_{\lam_1 \lam_2}^{SS_z}$ automatically regardless of
any kinds of vector mesons, i.e.
\be\label{ap6}
\sum_{\lam_1\lam_2}{\cal R}_{\lam_1 \lam_2}^{SS_z\dagger}{\cal R}_{\lam_1 \lam_2}^{SS_z}=1.
\ee
The explicit helicity components of ${\cal R}_{\lam_1 \lam_2}^{SS_z}$ for pseudoscalar and vector mesons
are given in the Appendix~\ref{sec:spin}.

For the radial wave function $\phi$, we use the same
Gaussian wave function for both pseudoscalar and vector mesons:
\be\label{rad}
\phi(x,{\bf k}_{\perp})=
\frac{4\pi^{3/4}}{\beta^{3/2}} \sqrt{\frac{\partial
k_z}{\partial x}} {\rm exp}(-{\vec k}^2/2\beta^2),
\ee
where $\beta$ is the variational parameter
fixed by the analysis of meson mass spectra~\cite{CJ_99}.
The longitudinal
component $k_z$ is defined by $k_z=(x-1/2)M_0 +
(m^2_2-m^2_1)/2M_0$, and the Jacobian of the variable transformation
$\{x,{\bf k}_\perp\}\to {\vec k}=({\bf k}_\perp, k_z)$ is given by
\be\label{jacob}
\frac{\partial k_z}{\partial x}
= \frac{M_0}{4 x (1-x)} \biggl\{ 1-
\biggl[\frac{m^2_1-m^2_2}{M^2_0}\biggr]^2\biggr\}.
\ee
The normalization of our wave function is then given by
\be\label{norm_phi}
\sum_{\lam_1\lam_2}\int\frac{dx d^2{\bf k}_\perp}{16\pi^3}
|\Psi^{SS_z}_{\lam_1\lam_2}(x,{\bf k}_{\perp})|^2
=
\int\frac{dx d^2{\bf k}_\perp}{16\pi^3}
|\phi(x,{\bf k}_{\perp})|^2.
\ee

Using the plus component of the currents and the longitudinal polarization vector,
the SLF calculation of the matrix element in Eq.~(\ref{eq:1}) is
\bea\label{ApSLF}
A^+_0 &=& \sqrt{N_c}\sum_{\lam_1\lam_2}\int\frac{dx d^2{\bf k}_\perp}{16\pi^3}
\phi(x,{\bf k}_\perp){\cal R}^{10}_{\lam_1\lam_2}(x,{\bf k}_\perp)
\frac{\bar{v}_{\lam_2}(p_2)}{\sqrt{p^+_2}}\gamma^+\frac{u_{\lam_1}(p_1)}{\sqrt{p^+_1}}
\nonumber\\
&=& f^{SLF}_V M \ep^+_{0}.
\eea
We again note that $\tilde{\ep}_0$ is used only in the calculation of the spin-orbit wave function
${\cal R}^{10}_{\lam_1\lam_2}(x,{\bf k}_\perp)$.
We then obtain the SLF result of the vector meson decay constant as follows (see Ref.~\cite{BL80} or Appendix A
for the Dirac matrix elements  for the helicity spinors in Eq.~(\ref{ApSLF}))~\cite{CJ_DA,Choi07,Jaus90,Hwang10}
\be\label{SLF_fv}
f^{SLF}_V = \frac{\sqrt{2N_c}}{{8\pi^3}}\int^1_0 dx \int d^2{\bf k}_\perp
\frac{\phi(x,{\bf k}_\perp)}{\sqrt{{\bf k}^2_\perp + {\cal A}^2}}
\left[ {\cal A} + \frac{ 2 {\bf k}^2_\perp}{D_{\rm LF}} \right].
\ee

\section{Correspondence between manifestly covariant model and LFQM}
\label{sec:cons}

In this section, we shall analyze the relations between
$f^{(h=0,1)}_V$ in the manifestly covariant BS model
and $f^{SLF}_V$ in the standard LFQM.
The main differences between the BS model and
the standard LFQM are attributed to the different spin structures of $q\bar{q}$
system (i.e. off-shellness vs. on-shellness) and the different vertex functions ($\chi$ vs. $\phi$).
In other words, while the results of the BS model allow the nonzero binding energy
$E_{\rm B.E.}=M^2-M^2_0$ but the SLF result is obtained from the zero binding energy
limit (i.e. $M\to M_0$). Thus one should take those different prescriptions into account
in connecting the two different models.

For the direct comparison between $f^{(h=0,1)}_V$ and $f^{SLF}_V$,
the LF covariant vertex function $\chi$  and the $D$ factor $D_{\rm con}=M + m_1 + m_2$ in $f^{(h=0,1)}_V$
may be replaced with the gaussian wave function $\phi$  and $D_{\rm LF}=M_0 + m_1 + m_2$ in $f^{SLF}_V$ via
\bea\label{eq:26}
{\rm Type\; I} : \;
 \sqrt{2N_c} \frac{ \chi(x,{\bf k}_\perp) } {1-x}
 &\to&  \frac{ \phi(x,{\bf k}_\perp) } {\sqrt{ {\cal A}^2 + {\bf k}^2_\perp }},
 \nonumber\\
 D_{\rm con} &\to& D_{\rm LF}.
 \eea
We denote Eq.~(\ref{eq:26}) as ``Type I" correspondence between the manifestly covariant BS model and
the standard LFQM. Essentially, Jaus~\cite{Jaus99} and subsequently the authors in~ \cite{Cheng04}
used this ``Type I" correspondence
when they connect the two different models and claimed that
$[f^{(h=1)}_V]_{\rm full}$ after  applying ``Type I" replacement is the correct
result in the LFQM
refuting effectively
the SLF result $f^{SLF}_V$.
However, one should note that this correspondence limits the replacement $M\to M_0$ only
in the $D$ factor so that the consistency of the replacement $M\to M_0$ within the LFQM is not assured.
This limitation leads to a consequence of not satisfying the symmetry constraint on DAs anticipated from the LFQM
as we discuss in the following section, Sec.~\ref{sec:Num}.

The correspondence between $\chi$ and $\phi$ given by Eq.~(\ref{eq:26})
has already been derived from the calculation of the zero-mode free
weak transition form factors between pseudoscalar and vector (or pseudoscalar)
mesons~\cite{Jaus99,Cheng04,CJ_Bc,CJ_PV}. We also explicitly
demonstrate in the Appendix~\ref{sec:psfv} that the correspondence between $\chi$ and $\phi$ can be obtained
from the comparison of the pseudoscalar meson decay constants between the two models.
On the other hand, the validity of the simple replacement $D_{\rm con} \to D_{\rm LF}$ for the $D$ factor
has not yet been clarified
as explicitly as
in the case of the vertex function replacement.
In the present work however, since we have now two exact forms of the decay constants
$[f^{(h=0)}_V]_{\rm full}$ and $[f^{(h=1)}_V]_{\rm full}$  obtained from the covariant BS model,
we are able to clearly check if ``Type I" correspondence is valid or not.
That is,  if  ``Type I" correspondence is correct, then
$[f^{(h=0)}_V]_{\rm full}$ and $[f^{(h=1)}_V]_{\rm full}$
should give the same result regardless of the SLF result $f^{SLF}_V$.
This is not the case as we shall show in Sec.~\ref{sec:Num}; e.g.,
$[f^{(h=1)}_V]_{\rm full}$ differs not only with $f^{SLF}_V$ but also with
$[f^{(h=0)}_V]_{\rm full}$ when ``Type I" replacement is used.

Considering that the result of $f^{SLF}_V$  is  essentially obtained from
the requirement of all constituents being on their respective mass shell,
we note that
it is more natural to apply the replacement $M\to M_0$
in each and every term including $M$ in the integrand of $f^{(h=0)}_V$ and $f^{(h=1)}_V$
than to apply it only in the $D$ factor as in the  case of  ``Type I" replacement.
For the self-consistency of the model,
we thus use the following replacement to connect the two different models:
\bea\label{eq:27}
{\rm Type\; II} : \;
 \sqrt{2N_c} \frac{ \chi(x,{\bf k}_\perp) } {1-x}
 &\to&  \frac{ \phi(x,{\bf k}_\perp) } {\sqrt{ {\cal A}^2 + {\bf k}^2_\perp }},
 \nonumber\\
 M &\to& M_0,
 \eea
in the integrand of the formulae for $f^{(h=0)}_V$ and $f^{(h=1)}_V$.
We denote Eq.~(\ref{eq:27}) as ``Type II" correspondence between the covariant BS model
and the standard LFQM. The essential point of this ``Type II" replacement is to apply the replacement of $M\to M_0$ to
all physical mass terms in the integrands of $f^{(h=0)}_V$ and $f^{(h=1)}_V$.
Without such self-consistent correspondence as given by ``Type II", it would not be possible to anticipate that
the LFQM with effective degrees of freedom represented by the constituent quark and antiquark
may satisfy the chiral symmetry consistent with the QCD.
With the self-consistent  ``Type II" replacement, we find numerically that the three different forms
$[f^{(h=0)}_V]_{\rm full}$, $[f^{(h=1)}_V]_{\rm full}$ and $f^{SLF}_V$ indeed yield the identical result.
Moreover, we find that the on-shell contribution $[f^{(h=1)}_V]_{\rm on}$
to $[f^{(h=1)}_V]_{\rm full}$ also gives the same result with the other three, i.e.
$[f^{(h=1)}_V]_{\rm on}=[f^{(h=1)}_V]_{\rm full}=[f^{(h=0)}_V]_{\rm full}=f^{SLF}_V$
in the standard LFQM.
From those observations, we conclude that the ``Type II" replacement
provides the self-consistent
correspondence in connecting the
covariant BS model and the standard LFQM.

Although those four different forms give the same result with each other when applying ``Type II" replacement,
their quark DAs are quite different.
Therefore, by checking the DAs as an important constraint of the model,
we are able to further pindown the self-consistent LF covariant forms of the decay constant.
The quark DA of a vector meson, $\phi_V(x,\mu)$, is the probability of finding collinear quarks up to
the scale $\mu$ in the $L_z=0$ (s-wave) projection of the meson wave function defined by
\be\label{eq:DA1}
\phi_V(x,\mu) = \int^{|{\bf k}_\perp|<\mu} \frac{ d^2{\bf k}_\perp}{16\pi^3}
\Psi^{SS_z}_{\lam_1\lam_2}(x,{\bf k}_{\perp}).
\ee
The dependence on the scale $\mu$ is
given by the QCD evolution~\cite{BL80} and can be calculated perturbatively.
However, the DAs at a certain low scale can be obtained by the necessary nonperturbative input
from LFQM. Moreover, the presence of the damping gaussian wave function allows us to perform
the integral up to infinity without loss of generality. The quark DA for a vector meson
is constrained by~\cite{CJ_DA,Hwang10}
\be\label{eq:DA}
\int^1_0\phi_V(x,\mu) dx = \frac{f_V}{2\sqrt{6}}.
\ee
One may also redefine the normalized quark DA as $\Phi_V(x)=(2\sqrt{6}/f_V)\phi_V(x)$ so that
$\int^1_0 dx \Phi_V(x)=1$.

For the equal quark and antiquark bound state meson such as $\rho$,
we find that only two forms of the decay constant, i.e.
$f^{SLF}_V$ from the longitudinal polarization and $[f^{(h=1)}_V]_{\rm on}$ from the transverse one,
yield the anticipated symmetric quark DA.  The other two forms, i.e.
$[f^{(h=1)}_V]_{\rm full}$ and $[f^{(h=0)}_V]_{\rm full}$,
that involve the corresponding zero-mode contributions
do not reproduce solely this fundamental constraint
expected from the symmetry associated with the two constituent masses $m_1$ and $m_2$
but reflect also the intrinsic characteristic of the zero-modes inherited from the vacuum property.
The involved zero-mode $Z_2$ (see Eq. (\ref{eq:20})) is apparently antisymmetric
under $x \leftrightarrow (1-x)$ when $m_1 = m_2$ is taken with the replacement $M \to M_0$ so that
the integration of $Z_2$ over $x$ vanishes as it reflects the vacuum property.
Consequently,
in the standard LFQM~\cite{CJ_DA,Choi07,Jaus90,Cheng97,Hwang10,Kon}, we
expect
two self-consistent LF covariant forms of the vector meson decay constant, i.e.
$f^{SLF}_V$ given by Eq.~(\ref{SLF_fv}) and $[f^{(h=1)}_V]_{\rm on}$ given by Eq.~(\ref{eq:23on}),
which provide the expected symmetric quark DAs for $m_1 = m_2$.
This expectation is realized by the ``Type II" replacement but not with the ``Type I" replacement.
The normalized quark DAs  obtained from $f^{SLF}_V$ and
 $[f^{(h=1)}_V]_{\rm on}$ (with ``Type II" replacement) correspond to the twist-2  $\phi^{||}_{2;V}(x)$
 and twist-3 $\phi^{\perp}_{3;V}(x)$, respectively.
 Our complete results for the twist-2 and twist-3 DAs in the standard LFQM are as follows:
\bea\label{T23DA}
 \phi^{||}_{2;V}(x) &=& \frac{2\sqrt{6}}{f_V}\int \frac{ d^2{\bf k}_\perp}{{16\pi^3}}
\frac{\phi(x,{\bf k}_\perp)}{\sqrt{{\bf k}^2_\perp + {\cal A}^2}}
\left[ {\cal A} + \frac{ 2 {\bf k}^2_\perp}{D_{\rm LF}} \right],
\nonumber\\
\phi^{\perp}_{3;V}(x) &=& \frac{2\sqrt{6}}{f_V}\int \frac{ d^2{\bf k}_\perp}{{16\pi^3}}
\frac{\phi(x,{\bf k}_\perp)}{\sqrt{{\bf k}^2_\perp + {\cal A}^2}}
\frac{1}{M_0}\biggl\{ \frac{{\bf k}^2_\perp + {\cal A}^2}{2x(1-x)} -{\bf k}^2_\perp
  + \frac{ (m_1 + m_2)}{D_{\rm LF}}{\bf k}^2_\perp
 \biggr\},
\eea
where we use $N_c=3$ and they satisfy the normalization given by Eq.~(\ref{DAap3}). We also note that while
 $f_V = f^{SLF}_V=[f^{(h=1)}_V]_{\rm on}$ (with ``Type II" replacement),  $f_V$ used in $\phi^{||}_{2;V}(x)$ and $\phi^{||}_{3;V}(x)$ correspond
 to  $f^{SLF}_V$ and $[f^{(h=1)}_V]_{\rm on}$, respectively.
In the next section, we show our findings explicitly with the numerical calculations of
the decay constants and quark DAs of vector mesons.

\section{Numerical Results}
\label{sec:Num}
In our numerical calculations within the standard LFQM, we use the set of the model parameters
(i.e. constituent quark masses and the gaussian parameters $\beta$) for the harmonic oscillator (HO) confining potentials
given in Table I, which was obtained from
the calculation of meson mass spectra using
the variational principle in our LFQM~\cite{CJ_99,CJ_DA,Choi07}.

%
\begin{table}[t]
\caption{The constituent quark mass (in GeV) and the gaussian parameters
$\beta$ (in GeV) for the HO potential obtained from the variational principle in our LFQM~\cite{CJ_99,CJ_DA,Choi07}. $q=u$ and $d$.}
\label{t1}
\begin{tabular}{cccccccc} \hline\hline
Model & $m_q$ & $m_s$ & $m_c$ &
$\beta_{qq}$ & $\beta_{qs}$ & $\beta_{qc}$
& $\beta_{cc}$  \\
\hline
HO & 0.25 & 0.48 & 1.8 & 0.3194 & 0.3419 &
0.4216 & 0.6998 \\
\hline\hline
\end{tabular}
\end{table}

\begin{table}[t]
\caption{Decay constants (in MeV) of vector mesons obtained
from $f^{(h=0)}_{\rm full}$,  $f^{(h=1)}_{\rm on}$, and $f^{(h=1)}_{\rm full}$ in the manifestly covariant
model but with Type I [Type II]
replacement compared with $f^{SLF}$ in our LFQM~\cite{CJ_99,CJ_DA,Choi07} and the experimental data~\cite{PDG12}. }
\label{t2}
\renewcommand{\tabcolsep}{0.6pc} 
\begin{tabular}{@{}ccccccc} \hline\hline
& &  $f^{(h=0)}_{\rm full}$ & $f^{(h=1)}_{\rm on}$ & $f^{(h=1)}_{\rm full}$ & $f^{SLF}$ & $f^{\rm exp.}$ \\
\hline
$\rho$& I & 256 & 299 & 299 &  - & -\\
      & II &  215 & 215 & 215 &  215 &
      220 (2)~\footnote{Exp. value for $\Gamma(\rho^0\to e^+ e^-)$.}, 209 (4)~\footnote{ Exp. value for $\Gamma(\tau\to\rho\nu_\tau)$.}\\
       \hline
$K^*$ & I &  272 & 320 & 320 & - & -\\
      & II &  223 & 223 & 223 & 223 & 217 (5)\\
       \hline
$D^*$ & I &  240 & 277 & 277 & - & -\\
      & II  &  212 & 212 & 212 & 212 & -\\
       \hline
$J/\psi$& I &  478 & 545 & 545 & - & -\\
        & II &  395 & 395 & 395 & 395 & 416 (6)\\
\hline\hline
\end{tabular}
\end{table}

In Table~\ref{t2}, we show the
results of the decay constants for  ($\rho, K^*, D^*, J/\psi$)  mesons obtained
from ($f^{(h=0)}_{\rm full}$,  $f^{(h=1)}_{\rm on}$,  $f^{(h=1)}_{\rm full}$) after applying  Type I   and Type II correspondences
and $f^{SLF}$  and compare them with the experimental
data~\cite{PDG12}.
For the Type I correspondence, the fact that the results of $f^{(h=0)}_{\rm full}$ are
different from those of $f^{(h=1)}_{\rm full}$ implies the breakdown of the covariance of the
decay constant.
For the Type II correspondence, however, the three different forms
$f^{(h=0)}_{\rm full}$, $f^{(h=1)}_{\rm full}$ and $f^{SLF}_V$  are found to yield the same numerical
results.
Although the on-shell contribution $f^{(h=1)}_{\rm on}$ to $f^{(h=1)}_{\rm full}$
gives identical result to the full solution for both Type I and II replacements,
it is quite obvious to use Type II correspondence in order to ensure the self-consistent covariant
description of a vector meson decay constant  in the standard LFQM.

%
\begin{figure}
\begin{center}
\includegraphics[height=7cm, width=7cm]{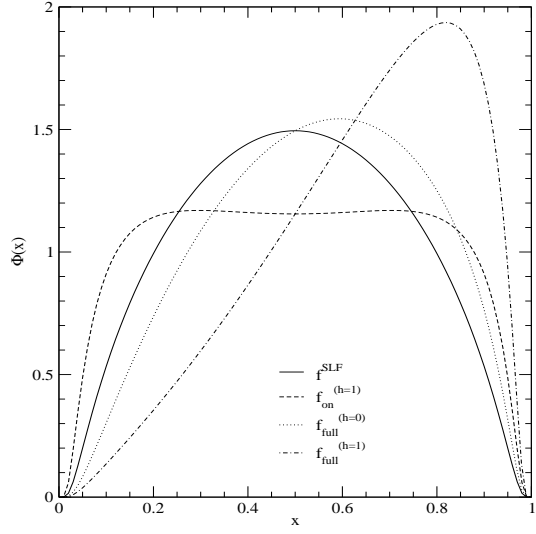}
\caption{\label{fig2} Normalized quark DAs $\Phi(x)$ of a rho meson obtained from the decay constants
$f^{SLF}$ (solid line), $f^{(h=1)}_{\rm on}$ (dashed line), $f^{(h=0)}_{\rm full}$ (dotted line),
and $f^{(h=1)}_{\rm full}$ (dot-dashed line), respectively. Only two of them, i.e. the results from
$f^{SLF}$ (solid line), $f^{(h=1)}_{\rm on}$ (dashed line), show the correct symmetric DAs.
}
\end{center}
\end{figure}

While four different forms, i.e. ($f^{(h=0)}_{\rm full},
f^{(h=1)}_{\rm on}, f^{(h=1)}_{\rm full}$) obtained from Type II correspondence and $f^{SLF}$, give
the identical results, they have different quark DAs as we discussed in the previous section, Sec.~\ref{sec:cons}.
Since $f^{(h=0)}_{\rm full}$ and $f^{(h=1)}_{\rm full}$ involve the corresponding zero-mode contributions,
they impose the intrinsic characteristic of the zero-modes, i.e. antisymmetric under $x \leftrightarrow (1-x)$,
inherited from the vacuum property. Thus,
the quark DAs from $f^{(h=0)}_{\rm full}$ and $f^{(h=1)}_{\rm full}$ do not satisfy the expected
constraint, i.e. symmetric DAs even for $m_1 = m_2$.
However, $f^{SLF}$ and $f^{(h=1)}_{\rm on}$ free from the explicit zero-mode contribution must yield
the symmetric DAs for the equal quark and antiquark bound state mesons such as $\rho$.

In Fig.~\ref{fig2}, we show the normalized quark DAs of a $\rho$ meson obtained from
the decay constants
$f^{SLF}$ (solid line), $f^{(h=1)}_{\rm on}$ (dashed line), $f^{(h=0)}_{\rm full}$ (dotted line),
and $f^{(h=1)}_{\rm full}$ (dot-dashed line), respectively.
As one can see from Fig.~\ref{fig2},  while two results  obtained from
$f^{SLF}$ and  $f^{(h=1)}_{\rm on}$  produce the anticipated symmetric DAs,
the other two results obtained from $f^{(h=0)}_{\rm full}$
and $f^{(h=1)}_{\rm full}$ show the asymmetric DAs reflecting the corresponding zero-mode contributions.

From the analysis of  the quark DAs for the $\rho$ meson, we
further pindown the correct covariant descriptions of vector meson decay constants.  That is,
we find that
$f^{(h=1)}_{\rm on}$ with Type II replacement
(but not with Type I replacement)
and $f^{SLF}$ provide self-consistent LF
covariant descriptions of  vector meson decay constants in the standard LFQM.
Since  $f^{SLF}$ [$f^{(h=1)}_{\rm on}$] is obtained from using ($J^+_W, \ep_{0}$)
[ ($J^\perp_W, \ep_{+}$) ], the normalized quark DAs  obtained from $f^{SLF}$ and
$f^{(h=1)}_{\rm on}$  correspond to the twist-2 DA $\phi^{||}_{2;V}(x)$ and
twist-3  DA  $\phi^{\perp}_{3;V}(x)$ given by  Eq.~(\ref{T23DA}), respectively.

\begin{figure}
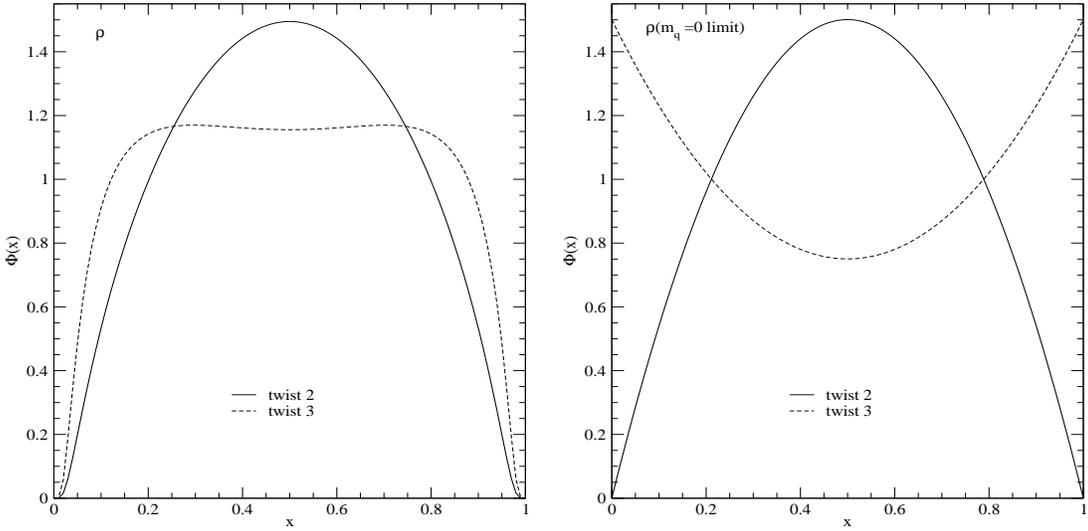

\begin{center}
\includegraphics[height=7cm, width=7cm]{Dfig3a.eps}
\hspace{0.2cm}
\includegraphics[height=7cm, width=7cm]{Dfig3b.eps}
\caption{\label{fig3} The twist-2 DAs $\phi^{||}_{2;V}(x)$ and twist-3
DAs $\phi^{\perp}_{3;V}(x)$ for $\rho$ meson with nonzero constituent quark masses
given in Table~\ref{t1} (left panel) compared to those with massless quark case (right panel)
}
\end{center}
\end{figure}
\begin{figure}
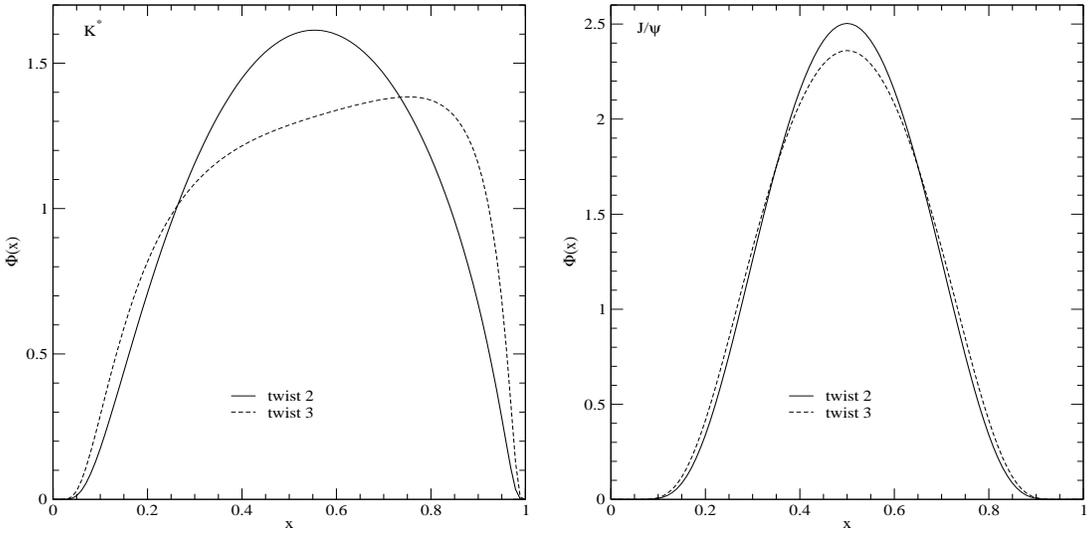

\vspace{0.8cm}
\begin{center}
\includegraphics[height=7cm, width=7cm]{Dfig4a.eps}
\hspace{0.2cm}
\includegraphics[height=7cm, width=7cm]{Dfig4b.eps}
\caption{\label{fig4} The twist-2 $\phi^{||}_{2;V}(x)$ and twist-3
DAs $\phi^{\perp}_{3;V}(x)$ for $K^*$ (left panel) and for $J/\psi$ (right panel) mesons.
}
\end{center}
\end{figure}

In Fig.~\ref{fig3}, we show the twist-2 DAs $\phi^{||}_{2;V}(x)$ (solid line) and twist-3
DAs $\phi^{\perp}_{3;V}(x)$ (dashed line) of the  $\rho$ meson in cases of  finite constituent quark masses
 (left panel)  and massless quark respecting
chiral
 symmetry (right panel).  We should note that our LFQM predictions of twist-2 and twist-3
  DAs in the
chiral
symmetry ($m_q\to 0$) limit remarkably reproduce the exact asymptotic DAs, i.e. $[\phi^{||}_{2;V}(x)]_{\rm as}=6 x (1-x)$
  and $[\phi^{\perp}_{3;V}(x)]_{\rm as}=(3/4)(1+\xi^2)$ where $\xi=2x -1$ anticipated from QCD sum rule predictions~\cite{Ball98}.
  This example may show again that the LFQM prediction
  satisfies the chiral symmetry consistent with the QCD if one correctly implement the zero-mode link to the QCD vacuum.
The quark mass correction is not large for the twist-2 $\phi^{||}_{2;V}(x)$,  however,
 it is very significant for the case of twist-3 $\phi^{\perp}_{3;V}(x)$ especially at the end point regions.
In Fig.~\ref{fig4}, we show $\phi^{||}_{2;V}(x)$ and  $\phi^{\perp}_{3;V}(x)$ for $K^*$ (left panel) and  $J/\psi$ (right panel) mesons.
For $K^*$ meson, we assign the momentum fractions $x$ for heavy $s$-quark and  $(1-x)$ for
the light $u(d)$-quark. One can see from the DAs of $\rho$ and $K^*$ that the SU(3) breaking effect
is more pronounced for $\phi^{\perp}_{3;V}(x)$ than for $\phi^{||}_{2;V}(x)$.
Comparing DAs of $\rho$ and $J/\psi$, we also find that the differences between $\phi^{||}_{2;V}(x)$
and $\phi^{\perp}_{3;V}(x)$ are significantly reduced for heavy quarkonium system and
$\phi^{||}_{2;V}(x)\simeq \phi^{\perp}_{3;V}(x)$ as the constituent quark mass $m_q \to \infty$.

For comparison purpose, we compute the expectation value of the longitudinal momentum, so called $\xi$-moments:
\be\label{eq:xi}
\la \xi^n\ra = \int^1_0 dx \xi^n \Phi(x),
\ee
where $\Phi=(\phi^{||}_{2;V}, \phi^{\perp}_{3;V})$. In Table~\ref{t3}, we summarize the first few $\xi$-moments
of twist-2 (T2) and twist-3 (T3) DAs for ($\rho, K^*, D^*, J/\psi$) mesons
obtained from our extracted DAs given by Eq.~(\ref{T23DA}). The numbers
in parenthesis for $\rho$ meson give the asymptotic values.

\begin{table}[t]
\caption{The $\xi$ moments of twist-2 (T2) and twist-3 (T3) DAs for ($\rho, K^*, D^*, J/\psi$) mesons. The numbers
in parenthesis for $\rho$ meson give the asymptotic values.}
\label{t3}
\renewcommand{\tabcolsep}{1pc} 
\begin{tabular}{@{}cccccc} \hline\hline
& &  $\la \xi^1\ra$ & $\la \xi^2\ra$ & $\la \xi^3\ra$ & $\la \xi^4\ra$  \\
\hline
$\rho$& T2 &  0  &  0.193 (0.200) & 0 &  0.078 (0.086) \\
      & T3 &  0 & 0.254 (0.400) & 0 &  0.120 (0.257) \\
       \hline
$K^*$ & T2 &  0.085 & 0.177 & 0.040 & 0.067 \\
      & T3 &  0.125 & 0.227 & 0.067 & 0.101 \\
       \hline
$D^*$ & T2 &  0.410 & 0.258 & 0.170 & 0.123 \\
      & T3 &  0.484 & 0.332 & 0.239 & 0.185 \\
       \hline
$J/\psi$& T2 &  0 &  0.083& 0 & 0.016 \\
        & T3 &  0 & 0.090 & 0 & 0.019 \\
\hline\hline
\end{tabular}
\end{table}

\section{Summary}
\label{sec:sum}

In this work, we extended our previous analysis~\cite{CJ_fv13} of the vector meson decay constant from the exactly
solvable manifestly covariant BS model to the more phenomenologically accessible realistic LFQM~\cite{CJ_99,CJ_DA,Choi07,Jaus90,Cheng97,Hwang10,Kon}. We discussed a
self-consistent covariant description of the vector meson decay constant in view of the
link between the chiral symmetry of QCD and the expected numerical results of the LFQM.
As the zero-mode contribution is locked into a single point of the LF longitudinal momentum in the meson decay process,
one of the constituents of the meson carries the entire momentum of the meson and it is important to capture the effect from
a pair creation of particles with zero LF longitudinal momenta from the strongly interacting vacuum.
The LFQM with effective degrees of freedom represented by the constituent quark and antiquark
may thus provide the view of effective zero-mode cloud around the quark and antiquark inside the meson.
Consequently, the constituents dressed by the zero-mode cloud may be expected to satisfy the chiral symmetry
of QCD. Our numerical results were consistent with this expectation and effectively indicated that
the constituent quark and antiquark in the standard LFQM~\cite{CJ_99,CJ_DA,Choi07,Jaus90,Cheng97,Hwang10,Kon}
could be considered as the dressed constituents including the zero-mode quantum fluctuations from the QCD vacuum.

As the SLF approach within the LFQM by itself is not amenable to determine the zero-mode contribution, we utilized the manifestly covariant model
to check the existence (or absence) of the zero-mode.
Performing a LF calculation in the covariant BS model, we computed the decay constants using two different combinations of
LF weak currents $J^\mu_W$ and polarization vectors $\epsilon_h$, i.e. $f^{(h=0)}_V$ obtained from
$(J^\mu_W,\epsilon_h)=(J^+_W,\epsilon_0)$ and $f^{(h=1)}_V$ from $(J^\perp_W,\epsilon_+)$,
 and checked the LF covariance of the decay constants.
We found in the manifestly covariant model that both combinations gave the same result with some
particular LF vertex functions if  the missing zero-mode contributions were properly taken into account.

We then substituted the radial and spin-orbit wave functions
with the phenomenologically accessible model wave functions provided by the LFQM and
compared $f^{(h=0,1)}_V$  obtained from the BS model with the decay constant $f^{SLF}_V$ obtained directly from the SLF
approach used in the LFQM~\cite{CJ_99,CJ_DA,Choi07,Jaus90,Cheng97,Hwang10,Kon}.
Linking the covariant BS model to the standard LFQM, we found the matching condition (i.e. ``Type II" correspondence)
between the two to give a self-consistent covariant description of the decay constant within the LFQM.
Using the ``Type II" correspondence, we were able to pin down two independent covariant forms of vector meson decay constants,
one obtained from $(J^+_W,\epsilon_0)$ and the other from $(J^\perp_W,\epsilon_+)$.
Although both of them yield the identical decay constant, each of them corresponds to different twist DA:
$(J^+_W,\epsilon_0)$ and $(J^\perp_W,\epsilon_+)$ correspond to twist-2 and twist-3 two-particle DAs, respectively.
Our twist-2 and twist-3 DAs not only satisfy  the fundamental constraint of the DAs anticipated from the isospin symmetry, i.e.
symmetric DAs for the equal quark and antiquark bound state mesons (e.g. $\rho$ meson), but also reproduce the correct
asymptotic DAs in the
chiral
symmetry limit.  Further analysis including the chirality odd and higher twist DAs is under consideration.

\acknowledgments
This work was supported by the Korean Research Foundation
Grant funded by the Korean Government (KRF-2010-0009019).
C.-R. Ji was supported in part by the US Department of Energy
(Grant No. DE-FG02-03ER41260).

\appendix

\section{Spin Structure in Standard Light-Front Quark Model}
\label{sec:spin}

It is instructive to use the appropriate basis of Dirac spinors:
\be\label{ap7}
u_{\lam}(p)
    = \frac{1}{\sqrt{p^+}}(/\!\!\!p + m)u_\lam,\;
    v_{\lam}(p)
    = \frac{1}{\sqrt{p^+}}(/\!\!\!p - m)v_\lam,
\ee
where
\be\label{ap8}
u_{\frac{1}{2}}\:
  =\left(
\begin{array}{c}
       1 \\
       0 \\
       0 \\
       0
\end{array}\; \right),
\; \; u_{-\frac{1}{2}}
  =\left(
     \begin{array}{c}
       0 \\
       0 \\
       0 \\
       1
     \end{array} \right),
\ee
and $v_\lam=u_{-\lam}$. In this basis the $\gamma$ matrices are
represented by
\be\label{ap9}
\gamma^0\: =
\: \left(
\begin{array}{cc}
0 & I \\
I & 0
\end{array} \; \right),
\;\gamma^i = \left(
\begin{array}{cc}
0 & \sigma^i \\
-\sigma^i & 0
\end{array} \; \; \right),
\ee
where $I$ is the $2\times 2$ unit matrix and $\sigma^{i}$ are
Pauli matrices. The normalization is
$\bar{u}_{\lam}(p)u_{\lam}(p)=-\bar{v}_{\lam}(p)v_{\lam}(p)=2m$.

We then obtain the spin-orbit wave functions of pseudoscalar  and vector mesons in the following
matrix forms:
\be\label{ap10}
{\cal R}^{00}_{\lam_1\lam_2}\:=
\frac{{\cal R}_0}{\sqrt{2}}
\left(
\begin{array}{cc}
        -k^L & {\cal A} \\
        -{\cal A} & -k^R
      \end{array}
    \right),\;
\ee
and
\bea\label{ap11a}
&&{\cal R}^{11}_{\lam_1\lam_2}\: =
{\cal R}_0
\left(
\begin{array}{cc}
{\cal A} +\frac{{\bf k}^2_\perp}{D_{\rm LF}} &
k^R\frac{xM_{0} + m_1}{D_{\rm LF}}\\
-k^R\frac{(1-x)M_0 + m_2}{D_{\rm LF}} &
-\frac{(k^R)^2}{D_{\rm LF}}
      \end{array}
    \right),\;
\nonumber\\
&&{\cal R}^{10}_{\lam_1\lam_2}\: =
\frac{{\cal R}_0}{\sqrt{2}}
\left(
\begin{array}{cc}
k^L\frac{{\cal M}}{D_{\rm LF}} &
{\cal A} + \frac{2{\bf k}^2_\perp}{D_{\rm LF}}\\
{\cal A} + \frac{2{\bf k}^2_\perp}{D_{\rm LF}} &
-k^R\frac{{\cal M}}{D_{\rm LF}}
      \end{array}
    \right),\;
\nonumber\\
&&{\cal R}^{1-1}_{\lam_1\lam_2}\: =
{\cal R}_0
\left(
\begin{array}{cc}
-\frac{(k^L)^2}{D_{\rm LF}} &
k^L\frac{(1-x)M_0 + m_2}{D_{\rm LF}} \\
-k^L\frac{xM_{0} + m_1}{D_{\rm LF}} &
{\cal A} +\frac{{\bf k}^2_\perp}{D_{\rm LF}}
      \end{array}
    \right),\;
\nonumber\\
\eea
where $k^{R(L)}=k^x \pm ik^y$,
${\cal R}_0=\frac{1}{\sqrt{{\cal A}^2 +{\bf k}^2_\perp}}$, and
${\cal M}=(1-2x)M_0 + m_2 - m_1$ .

\section{Pseudoscalar Meson Decay Constant}
\label{sec:psfv}

The decay constant $f_{P_s}$ of a pseudoscalar meson with the four-momentum $P$ and mass $M$
as a $q{\bar q}$ bound state
is defined by the matrix element of the axial vector current

\be\label{eq:b1}
\la 0|{\bar q}\gamma^\mu\gamma_5 q|P_s(P)\ra
= if_{P_s} P^\mu.
\ee
The matrix element $B^\mu \equiv \la 0|{\bar q}\gamma^\mu\gamma_5 q|P_s(P)\ra$ is given in
the one-loop approximation as a momentum integral
\be\label{eq:b2}
B^\mu = N_c
\int\frac{d^4k}{(2\pi)^4} \frac{H_V} {N_p N_k} S^\mu_{P_s},
\ee
where
\be\label{eq:b3}
 S^\mu_{P_s}  =  {\rm Tr}\left[\gamma^\mu\gamma_5\left(\slash \!\!\!p+m_1 \right)
 \gamma_5
 \left(-\slash \!\!\!k + m \right) \right],
\ee
which can be separated into the
on-mass-shell propagating part $[S^\mu_{P_s}]^{\rm on}$ and
the off-mass-shell instantaneous part $[S^\mu_{P_s}]^{\rm inst}$ as
\be\label{eq:b4}
[S^\mu_{P_s}]^{\rm on}=
{\rm Tr}\left[\gamma^\mu\gamma_5\left(\slash \!\!\!p_{\rm on} + m_1 \right)
 \gamma_5
 \left(-\slash \!\!\!k_{\rm on} + m \right) \right],
\ee
and
\bea\label{eq:b5}
[S^\mu_{P_s}]^{\rm inst} &=& -\frac{1}{2}(\Delta_k^-)
{\rm Tr}\left[\gamma^\mu\left(-\slash \!\!\!p_{\rm on} + m_1 \right)
 \gamma^+\right]
 -\frac{1}{2}(\Delta_p^-)
{\rm Tr}\left[\gamma^\mu\gamma^+\left(-\slash \!\!\!k_{\rm on} + m_2 \right)
\right].
\eea
For pseudoscalar meson decay constant, we find that while the plus component of the currents
is immune to the zero-mode contribution, the minus component of the current receives the zero mode.

Explicitly, the pseudoscalar decay constant
$[f_{P_s}]_{\rm full}=[f_{P_s}]_{\rm val}$
obtained from $\mu=+$ is
\bea\label{eq:b6}
[f_{P_s}]_{\rm full}
&=& \frac{N_c}{4\pi^3}\int^1_0\frac{dx}{(1-x)}\int d^2{\bf k}_\perp
\chi(x,{\bf k}_\perp) {\cal A}.
\eea
For $\mu=-$, while the valence contribution (i.e. $k^-=k^-_{\rm on}$)
to the trace term is given by
\bea\label{eq:b7}
[S^-_{P_s}]^{\rm val} &=& [S^-_{P_s}]^{\rm on} + [S^-_{P_s}]^{\rm inst}
= 4
\left[ m_2 M^2 + (m_1 - m_2)\frac{{\bf k}^2_\perp + m^2_2}{1-x} \right].
\nonumber\\
\eea
we find the singular term in $S^-_{P_s}$ in the limit of $x\to 0$ when $p^-=p^-_{\rm on}$
as
\be\label{eq:b8}
\lim_{x\to 0} S^-_{P_s}(p^-=p^-_{\rm on})= 4 (m_2-m_1) p^-.
\ee
Thus, the corresponding zero-mode operator is given by
\be\label{eq:b9}
[S^-_{P_s}]^{\rm Z.M.} = 4 (m_2-m_1) (-Z_2).
\ee
Adding $[S^-_{P_s}]^{\rm val}$ and $[S^-_{P_s}]^{\rm Z.M.}$, we obtain
\bea\label{eq:b10}
[S^-_{P_s}]^{\rm tot}= [S^-_{P_s}]^{\rm val} + [S^-_{P_s}]^{\rm Z.M.}
= 4 M^2 {\cal A},
\eea
and the full solution obtained from $\mu=-$ is shown to be completely equal to
the one given by Eq.~(\ref{eq:b6}).

On the other hand, the SLF result of a pseudoscalar meson decay constant is given
by
\be\label{eq:b11}
f^{SLF}_{P_s} = \frac{\sqrt{2N_c}}{{8\pi^3}}\int^1_0 dx \int d^2{\bf k}_\perp
\frac{\phi(x,{\bf k}_\perp)}{\sqrt{{\bf k}^2_\perp + {\cal A}^2}}
{\cal A}.
\ee
Comparing Eqs.~(\ref{eq:b6}) and~(\ref{eq:b11}), we find that
the vertex function $\chi$ in Eq.~(\ref{eq:b6}) may be replaced with the gaussian wave
function $\phi$ in Eq.~(\ref{eq:b11}) via
\begin{eqnarray*}
 \sqrt{2N_c} \frac{ \chi(x,{\bf k}_\perp) } {1-x}
 &=& \frac{ \phi(x,{\bf k}_\perp) } {\sqrt{ {\cal A}^2 + {\bf k}^2_\perp }}.
 \end{eqnarray*}

\end{document}